\documentstyle[12pt]{article}

\begin{document}
\begin{flushright}
SNUTP-98-102
\end{flushright}
\vspace{-11mm}
\begin{flushright}
WU-AP/73/98
\end{flushright}
\vspace{-11mm}
\begin{flushright}
YITP-98-64
\end{flushright}
\vspace{-11mm}
\begin{flushright}
hep-ph/9901436
\end{flushright}

\vspace{5mm}

\begin{center}
{\Large \bf Gauged Monopole-Bubble}\\[10mm]
Yoonbai Kim$^{\rm{a},}$\footnote{E-mail: yoonbai$@$cosmos.skku.ac.kr},
Seung Joo Lee$^{\rm{a},}$\footnote{E-mail: suphy$@$newton.skku.ac.kr},
Kei-ichi Maeda$^{\rm{b},}$\footnote{E-mail: maeda$@$mse.waseda.ac.jp},
Nobuyuki Sakai$^{\rm{c},}$\footnote{E-mail: sakai$@$yukawa.kyoto-u.ac.jp}
\\ \em
$^{\rm{a}}$Department of Physics and Institute of Basic Science,
Sungkyunkwan University,\\
Suwon 440-746, Korea\\
$^{\rm{b}}$Department of Physics, Waseda University,
Tokyo 169-8555, Japan\\
$^{\rm{c}}$Yukawa Institute for Theoretical Physics, Kyoto University,
Kyoto 606-8502, Japan
\end{center}

\vspace{5mm}

\begin{abstract}
The decay of a metastable false vacuum by bubble nucleation is studied in the
high temperature limit of the gauge theory in which an SO(3) gauge symmetry 
is spontaneously broken to an SO(2). The effects of internal symmetry 
are so drastic that, in addition to the known Euclidean bounce solution, 
there exists a new bubble solution involving a 't Hooft-Polyakov monopole 
at its center the moment it is nucleated. The decay rate and evolution are analyzed.
\end{abstract}



\newpage
\baselineskip 18pt

\setcounter{equation}{0}
\section{Introduction}
Since the Euclidean bounce solution of a scalar field theory has been
introduced for a description of first-order phase transition, the scalar
sector of a given model determines mainly the metastable decay
process~\cite{Col,CC}. For cosmological phase transitions in the early
universe, two more ingredients are needed to be included: one is
gravitation~\cite{CD} and the other may be  temperature~\cite{Lin}. When a
spontaneous continuous symmetry breaking is taken into account, an enhancement
of the tunneling is expected~\cite{Kus}. However, the uniqueness of the bounce
O(3) symmetric solution as a nucleated bubble~\cite{Col2} has never been doubted
before three of the authors reported the global {\it monopole-bubble} solution in 
the model of a scalar triplet with global SO(3) symmetry~\cite{Kim,KMS}.

In this paper, we will address the same question to a gauge theory of SO(3)
symmetry, and show that there exists another {\it gauged monopole-bubble}
solution in which the center includes a 't Hooft-Polyakov monopole~\cite{HP}.
The nucleation rate of gauged monopole-bubbles is lower than that of the
bounce, but it is higher than that of global monopole-bubbles. Particularly,
when the mass of monopole becomes small in the strong gauged coupling limit, the
decay channel through a gauged monopole-bubble with thick wall is quite 
considerable in the limit of high temperature. If the size of a nucleated 
gauged monopole-bubble is smaller than its critical size, the
bubble wall starts to shrink and then the monopole at its center disappears.
On the other hand, for the opposite case, the bubble wall grows and the 
monopole survives safely.

The remainder of this paper is organized as follows. In Sec. II we demonstrate
the existence of a new gauged monopole-bubble solution in addition to the known
bounce and the global monopole-bubble solution. In Sec. III we compute the
nucleation rates of those solutions and present the evolution of bubbles.
Section IV contains some brief concluding remarks.

\section{Gauged Monopole-bubble Solution}
The Euclidean action of Georgi-Glashow model with an SO(3) gauge 
symmetry is
\begin{eqnarray}\label{action}
S=\int^{\beta}_{0}dt_{E}\int 
d^{3}x\biggl\{\frac{1}{4}F^{a\;2}_{\mu\nu}
+\frac{1}{2}(D_{\mu}\phi^{a})^{2}+V\biggr\},
\end{eqnarray}
where the field strength tensor is $F^{a}_{\mu\nu}\equiv\partial_\mu A^{a}_{\nu} -
\partial_\nu A^{a}_{\mu}+ e\epsilon^{abc}A^{b}_{\mu}A^{c}_{\nu}$, and the
covariant derivative is ${D_{\mu} \phi}^{a}\equiv\partial_\mu\phi^{a} +
e\epsilon^{abc}A^{b}_{\mu} \phi^{c}$.
In order to describe a first-order phase transition, we choose a 
$\phi^{6}$-potential:
\begin{eqnarray}\label{pot}
V(\phi)=\frac{\lambda}{v^{2}}(\phi^{2}+\alpha v^{2})(\phi^{2}
-v^{2})^{2},
\end{eqnarray}
where the scalar amplitude $\phi$ is defined by 
$\phi=\sqrt{\phi^{a}\phi^{a}}$
$(a=1,2,3)$, and a parameter $\alpha$ $\;(0<\alpha<1/2)$ governs the 
transition
rate from the symmetric 
false vacuum at $\phi=0$ to a broken true vacuum at $\phi=v$.

In the high temperature limit, i.e., $\beta\rightarrow 0$, a static
electrically-neutral object satisfies
\begin{eqnarray}\label{seq1}
(D_{i}^{2}\phi)^{a}=\frac{\partial V}{\partial\phi^{a}}\;\;,
\end{eqnarray}
\begin{eqnarray}\label{seq2}
(D_{j}F_{ij})^{a}=e\epsilon_{abc}\phi^{b}(D_{i}\phi)^{c}.
\end{eqnarray}
The ordinary bounce configuration is obtained under the ansatz of the scalar 
field $\phi^{a}=(0,0,f(r))$ with boundary conditions,
$df/dr|_{r=0}=0$ and $f(r=\infty)=0$. Here one may ask a question whether 
or not the gauge field affects the bounce configuration. Since the 
SO(3)
gauge symmetry is spontaneously broken to an SO(2) symmetry and the 
first two
components of the scalar field vanish, the gauge field decouples from 
the bounce solution. Therefore, even under a general ansatz for the
electrically-neutral gauge field~\cite{Wit}, it is easy to show 
decoupling of
the gauge field from the bounce solution~\cite{KKMS}.

\begin{figure}

\setlength{\unitlength}{0.1bp}
\begin{picture}(3600,2160)(0,0)
\put(1850,150){\makebox(0,0){\Large${vr}$}}
\put(3550,1230){%
\makebox(0,0)[b]{\Large\shortstack{${K}$}}%
}
\put(100,1230){%
\makebox(0,0)[b]{\Large\shortstack{${{\phi}/{v}}$}}%
}
\put(3300,2060){\makebox(0,0)[l]{1}}
\put(3300,1728){\makebox(0,0)[l]{0.8}}
\put(3300,1396){\makebox(0,0)[l]{0.6}}
\put(3300,1064){\makebox(0,0)[l]{0.4}}
\put(3300,732){\makebox(0,0)[l]{0.2}}
\put(3300,400){\makebox(0,0)[l]{0}}
\put(3253,300){\makebox(0,0){10}}
\put(2692,300){\makebox(0,0){8}}
\put(2132,300){\makebox(0,0){6}}
\put(1571,300){\makebox(0,0){4}}
\put(1011,300){\makebox(0,0){2}}
\put(450,300){\makebox(0,0){0}}
\put(400,2060){\makebox(0,0)[r]{1}}
\put(400,1728){\makebox(0,0)[r]{0.8}}
\put(400,1396){\makebox(0,0)[r]{0.6}}
\put(400,1064){\makebox(0,0)[r]{0.4}}
\put(400,732){\makebox(0,0)[r]{0.2}}
\put(400,400){\makebox(0,0)[r]{0}}
\end{picture}
\vspace{3mm}
\caption{Plots of monopole-bubbles for $\lambda=1$ and $e=0.3$.
The dashed lines correspond to the scalar amplitude $\phi(r)/v$ and the
gauge field $K(r)$ of a thin-wall monopole-bubble ($\alpha$ = 0.15),
and the solid lines to those of a thick-wall monopole-bubble 
($\alpha$ = 0.35).}
\label{fig1}
\end{figure}
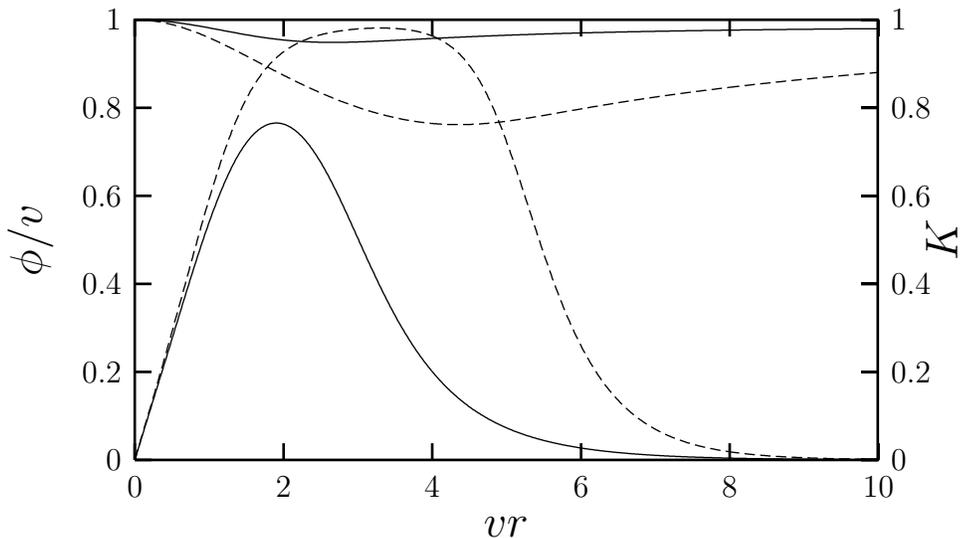

In this paper we are interested in a different bubble solution which 
is obtainable under the hedgehog ansatz:
\begin{eqnarray}\label{an1}
\phi^{a}=\hat{r}^{a}\phi(r),~~~A^{a}_{i}=
\epsilon^{aij}\hat{r}^{j}\frac{1-K(r)}{er}.
\end{eqnarray}
Substituting Eq.~(\ref{an1}) into Eqs.~(\ref{seq1}) and (\ref{seq2}), 
we have
\begin{eqnarray}\label{meq1}
\frac{d^{2}\phi}{dr^{2}}+\frac{2}{r}\frac{d\phi}{dr}
-2\frac{K^{2}}{r^{2}}\phi=\frac{dV}{d\phi}\;,
\end{eqnarray}
\begin{eqnarray}\label{meq2}
\frac{d^{2}K}{dr^{2}}=K\Bigl(\frac{K^{2}-1}{r^{2}}+e^{2}\phi^{2}\Bigr)\;.
\end{eqnarray}
Note that the ordinary bounce solution $\phi_{\rm{b}}^{a}$ 
corresponds to $K=0$
solution where a Dirac monopole in Eq.~(\ref{an1}) decouples from this
solution~\cite{Col}, and that the global monopole-bubble 
$\phi_{\rm{gm}}^{a}$ is
supported by the replacement from $K(r)^{2}$ in Eq.~(\ref{meq1}) to 1
only when the gauge coupling $e$ vanishes because of Eq.~(\ref{meq2})
~\cite{Kim,KMS}.
The first-order transition from a symmetric false
vacuum ($\phi=0$) to a true broken vacuum ($\phi=v$) when
$0<\alpha<1/2$; possible boundary conditions are read as follows: 
regularity of the fields at the origin gives
\begin{eqnarray}\label{bc0}
\phi(r=0)=0,~~~K(r=0)=1,
\end{eqnarray}
and the spatial infinity should be in the initial setting before 
transition:
\begin{eqnarray}\label{bcinf}
\phi(r\rightarrow\infty)=0,~~~K(r\rightarrow\infty)=1.
\end{eqnarray}
Of course, the trivial solution of false symmetric vacuum, i.e.,
$\phi=0$ and $K=1$, is an unwanted one.
Therefore, as shown in Fig.~\ref{fig1},
the nontrivial bubble solution of our interest is: $\phi(r)$ 
increases from $\phi=0$ at the beginning, reaches a maximum value 
$\phi_{\rm max}$, and decreases to 0 as $r$ goes to infinity. 
On the other hand, $K(r)$ starts to decrease from 1, arrives at a minimum, and then 
grows up to 1 at spatial infinity.

\begin{figure}
\setlength{\unitlength}{0.1bp}
\begin{picture}(3600,2160)(0,0)
\put(2000,150){\makebox(0,0){\Large${vr}$}}
\put(100,1230){%
\makebox(0,0)[b]{\Large\shortstack{$T^{0}_{0}/e^{2}v^{4}$}}%
}
\put(3550,300){\makebox(0,0){8}}
\put(2775,300){\makebox(0,0){6}}
\put(2000,300){\makebox(0,0){4}}
\put(1225,300){\makebox(0,0){2}}
\put(450,300){\makebox(0,0){0}}
\put(400,2060){\makebox(0,0)[r]{1.6}}
\put(400,1645){\makebox(0,0)[r]{1.2}}
\put(400,1230){\makebox(0,0)[r]{0.8}}
\put(400,815){\makebox(0,0)[r]{0.4}}
\put(400,400){\makebox(0,0)[r]{0}}
\end{picture}
\vspace{3mm}
\caption{Profiles of the energy density for $\lambda=1$ and 
$\alpha=0.35$. 
The solid line stands for a monopole-bubble with a monopole wall 
($e=0.1$),
and the dashed line for a monopole-bubble without a monopole wall 
($e=0.65$).}
\label{fig2}
\end{figure}
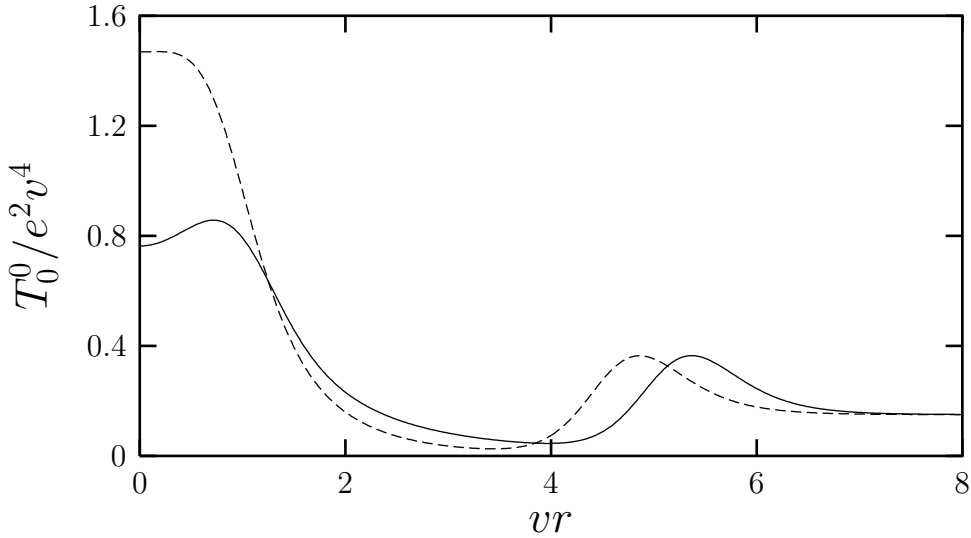

To read the physics of the obtained sphaleron-type monopole-bubble 
solution,
let us examine energy density:
\begin{eqnarray}
T^{0}_{\;0}=\frac{1}{e^{2}r^{2}}\biggl[\biggl(\frac{dK}{dr}\biggl)^{2}+
\frac{(K^{2}-1)^{2}}{2r^{2}}\biggr]+\frac{1}{2}\biggl(\frac{d\phi}{dr}
\biggr)^{2}+\frac{K^{2}}{r^{2}}\phi^{2}+V. \label{ener}
\end{eqnarray}
For small $r$ we attempt a power series solution:
\begin{eqnarray}\label{pr0}
\phi(r)\approx\phi_{0}\biggl[r-\frac{1}{10}(4\kappa_{0}
-m^{2})r^{3}+\cdots\biggr],
\end{eqnarray}
which is formally odd under $r\rightarrow -r$, and
\begin{eqnarray}\label{kr0}
K(r)\approx 1-\kappa_{0}r^{2}+\frac{1}{10}(3\kappa^{2}_{0}+
e^{2}\phi_{0}^{2})r^{4}+\cdots,
\end{eqnarray}
which is even in form under $r\rightarrow -r$. Here $m$ denotes the 
mass of scalar particles at the symmetric vacuum, i.e.,
$m=\sqrt{d^{2}V/d\phi^{2}|_{\phi=0}}=\sqrt{2\lambda(1-2\alpha)}v$.
If $\kappa_{0}$ is negative, then Eq.~(\ref{meq2})
says that $K(r)$ is a monotonic increasing function bounded below 
1. It means that any solution with a negative $\kappa_{0}$ cannot 
satisfy the boundary condition (\ref{bcinf}) at spatial infinity, and 
thereby $\kappa_{0}$ must be positive.
A characteristic of this new monopole-bubble,
which makes it distinguished from the ordinary bounce configuration, 
is to have a matter lump at the center of the bubble.
If we look into the behavior of energy density for small $r$ by 
use of the formulas (\ref{pr0}) and (\ref{kr0}), it has
\begin{eqnarray}\label{ener2}
T^{0}_{\;0}&=& 
6\kappa_{0}^{2}{e^{2}}+{3\phi_{0}^{2}\over 2}+\lambda\alpha v^4
+\Biggr[m^{2}\phi_{0}^{2}
-2\kappa_{0}\biggr(\frac{4\kappa_{0}^2}{e^{2}}+3\phi_{0}^{2}\biggr)
\Biggr]r^{2}+\cdots.
\end{eqnarray}
The first term in Eq.~(\ref{ener2}) is always positive, which stems 
from the winding between the 3-dimensional space and the SO(3) internal space.
It is an evidence of the formation of an extended object
at the center of the bubble. The behavior of energy density at the 
center depends on the sign of the second term in Eq.~(\ref{ener2}). If 
it is positive, $T^{0}_{\;0}(r)$ increases from 0, reaches a maximum, 
and decreases, as shown in the solid line of Fig.~\ref{fig2}.
This structure is like a domain wall with the width 
$1/m_{\rm H}\sim 1/\sqrt{8\lambda(1+\alpha)} v$.
In the case of global monopole-bubbles, the existence of this wall is 
automatic since $\kappa_{0}$ is zero and the second term is always 
positive \cite{Kim,KMS}. However, when the gauge coupling is 
sufficiently large so as to make the second term negative, strong gauge 
repulsion can sweep the monopole wall
away and the energy density is a decreasing function near the origin, 
as shown in the dashed line of Fig.~\ref{fig2}.

At the asymptotic region for large $r$, the scalar field approaches
the boundary value in Eq.~(\ref{bcinf}) exponentially,
\begin{eqnarray}\label{pinf}
\phi\approx\phi_{\infty}\frac{1+mr}{(mr)^{2}}e^{-mr}.
\end{eqnarray}
The gauge field has a long range power tail because the gauge 
boson becomes massless in the symmetric phase,
\begin{eqnarray}\label{kinf}
K(r)\approx 1-\frac{\kappa_{\infty}}{r}+\frac{3\kappa^{2}_{\infty}}
{4r^{2}}+\cdots,
\end{eqnarray}
where $\kappa_{\infty}$ should also be nonnegative for the same reason
as $\kappa_{0}$ is in Eq.~(\ref{kr0}). Outside the bubble wall, the vacuum is
in the symmetric phase with $\big<\phi\big>=0$, and the SO(3) gauge 
field is massless. Specifically, the magnetic field looks like that of a 
dipole:
\begin{eqnarray}
F^{a}_{ij}\sim\frac{1}{e}(\epsilon^{ajk}\hat{r}^{i}-\epsilon^{aik}\hat{r}^{j}
)\frac{\hat{r}^{k}}{r}\frac{dK}{dr}\sim {\cal O} (1/r^{3}).
\end{eqnarray}
The corresponding leading term of the energy density, 
$T^{0}_{\;0}-V(0)$, in
Eq.~(\ref{ener}) is of order $1/r^{6}$,
and this fast decay can also be read from Fig.~\ref{fig2}.
This means that the monopole at the center is screened by the bubble wall.

In the remainder of this section we make analytic arguments on
the properties of gauged monopole-bubbles under the thin-wall assumption. 
If the energy gap between the false vacuum and the true vacuum,
$\Delta V\equiv V(0)-V(v)=\lambda\alpha v^{4}$, is small enough, the 
bubble radius becomes
much larger than both of the monopole size and the width of bubble
wall, and we may adopt the thin-wall approximation.
The typical scale of the monopole core or the outer wall is given
$\sim1/m_{\rm H}$ by the 
configuration of the Higgs field, while it is given $\sim 1/ev$ by the 
configuration of the gauge field. Here we assume that both scales are 
much smaller than the bubble radius, and therefore we implicitly 
exclude the weak coupling case $e\ll1$.
First, let us consider the behavior of the fields near the maximum scalar amplitude
$\phi_{\rm max}$, which is close to $v$ (see the solid lines in Fig.~\ref{fig1}).
Around $\phi=\phi_{\rm max}$, the derivative of the scalar field 
${d\phi}/{dr}$ is close to zero, and the slope of the potential 
${dV}/{d\phi}$ is almost
flat. Therefore, if the corresponding value of $K$ at the turning point
$r_{\rm turn}$ with $\phi(r_{\rm turn})=\phi_{\rm max}$ is negligibly 
small, which will turn out to be indeed the case, the scalar field 
equation (\ref{meq1}) implies that the quadratic derivative of 
the potential $d^{2}V/d\phi^{2}$ is also very small near $\phi_{\rm max}$.
Thus we see that $\phi$ stays near $\phi_{\rm max}$ for a long range
of $r$. In this case, the first term of the
right-hand side of Eq.~(\ref{meq2}) becomes negligible near 
$\phi\approx\phi_{\rm max}$ since $r_{\rm turn}$ is large enough.
Therefore, $K$ falls to zero exponentially, i.e., $K\sim 
e^{-e\phi_{\rm max}r}$, where the scale $e\phi_{\rm max}\approx ev$ in 
the exponential
is nothing but the mass of gauge boson in the Higgs phase. We can 
also confirm
that $\phi$ approaches exponentially to the true vacuum expectation value
$v$, i.e., $\phi\approx v-\phi_{\rm max}{e^{-m_{\rm H}r}}/{m_{\rm H}r}$
where $\phi_{\rm turn}$ is a constant to be 
fixed by boundary conditions (\ref{bc0}) and (\ref{bcinf}).

Secondly, we discuss how energy density is distributed in the region between 
the monopole core and the bubble wall. Because this region is in a 
symmetry-broken phase of which the breaking
pattern is SO(3)$\rightarrow$SO(2), one of the three internal gauge degrees
remains massless and it is to be identified as the photon field
described by
\begin{eqnarray}\label{abel}
F_{\mu\nu}=\frac{\phi^{a}}{\phi}F^{a}_{\mu\nu}+\frac{1}{e\phi^{3}}
\epsilon_{abc}\phi^{a}(D_{\mu}\phi)^{b}(D_{\nu}\phi)^{c}.
\end{eqnarray}
Inserting the monopole ansatz (\ref{an1}) into Eq.~(\ref{abel}), we 
obtain zero electric field but
in contrast the magnetic field with the monopole charge
$g=1/e$, i.e., $F_{ij}=\epsilon^{ija}\hat{r}^{a}/er^{2}$.
Since $\phi\approx v$ and $K\approx 0$ around $r=r_{\rm turn}$,
one can easily notice that the contribution of the
magnetic field comes from the first term of Eq.~(\ref{abel}), and 
then identify the matter lump produced at the center as a 
't Hooft-Polyakov monopole \cite{HP}. If we look at the expression of energy 
density (\ref{ener}) near $\phi=\phi_{\rm max}\sim v$,
\begin{eqnarray}\label{mener}
T^{0}_{\;0}\sim\frac{1}{2 e^{2}r^{4}}+V(v),
\end{eqnarray}
the above leading term is interpreted as magnetic energy of a 't 
Hooft-Polyakov monopole, and
its total energy is finite when we do not take into 
account the contribution from latent heat term $V(v)$. This is different
from the global monopole-bubble case, where the produced global monopole
contributes a $1/r^{2}$ term to energy density, since this scalar phase
term, which would make the energy diverge, is eaten up by the gauge field.

Finally, we argue how the gauge field affects the bubble size. Let us
estimate the radius by Coleman's action-minimum method \cite{Col,Lin2}. 
After assuming spherical symmetry and the hedgehog configuration
(\ref{an1}), the action (\ref{action}) is reduced to
\begin{equation}\label{action2}
S=4\pi\beta\int^{\infty}_{0}r^2dr
\left\{\frac12{\phi'}^2+{K^2\phi^2\over r^2}+V
+{(1-K^2)^2\over 2e^2r^4}+{{K'}^2\over e^2r^2}\right\}.
\end{equation}
By the thin-wall assumption, the core is approximately described
as a 't Hooft-Polyakov monopole, and we may approximate $\phi=v$ and 
$K=0$ between the monopole core and the bubble wall.
Then the difference between the action for a gauged monopole-bubble  
and that for the false vacuum is
\begin{eqnarray}
B_{\rm lm}&\equiv&S(\phi)-S(\phi=0) \nonumber\\
&=&\beta\left(-{4\pi\over 3}R^3\Delta V+4\pi\sigma_{\phi}R^2+M_K
-{2\pi\over e^2R}\right)+{\rm constant~core~term},
\end{eqnarray}
where $R$ is the bubble radius, and $\sigma_{\phi}$ and $M_L$ are 
constants, which are defined as
\begin{equation}
\sigma\equiv\frac12\int^{R+\epsilon}_{R-\epsilon}dr\left({d\phi\over dr}\right)^2,
~~~ M_K\equiv{4\pi\over e^2}\int^{R+\epsilon}_{R-\epsilon}dr
             \left({dK\over dr}\right)^2t.
\end{equation}
Similarly, for a normal bubble and a global monopole-bubble, we obtain
\begin{eqnarray}
B_{\rm b}&=&\beta\left(-{4\pi\over 3}R^3\Delta V+4\pi\sigma_{\phi}R^2\right),
\\
B_{\rm gm}&=&\beta\left(-{4\pi\over 3}R^3\Delta V
+4\pi\sigma_{\phi}R^2+4\pi v^2R\right)+{\rm constant~core~term}.
\end{eqnarray}
Each radius is determined by the condition $dB/dR=0$:
\begin{equation}
R_{\rm b}={2\sigma_{\phi}\over\Delta V}, ~~~
R_{\rm gm}\approx R_{\rm b}+{v^2\over 2\sigma_{\phi}}
\end{equation}
What we need is an inequality among $R_{\rm b},~ R_{\rm gm},~ R_{\rm 
lm}$ rather than a complicated expression of $R_{\rm lm}$. Thus we 
evaluate
\begin{equation}\label{dB}
{dB_{\rm lm}\over dR}\Big|_{R=R_{\rm b}}
={2\pi\beta\over e^2R_{\rm b}^2}>0, ~~~
{dB_{\rm lm}\over dR}\Big|_{R=R_{\rm gm}}
=2\pi\beta\left({1\over 4e^2R_{\rm gm}^2}-v^2\right)<0,
\end{equation}
where the last inequality is supported by the initial assumption 
$R\gg1/ev$. Because $R_{\rm lm}$ is a local maximum of $B_{\rm lm}$, 
Eq.~(\ref{dB}) imply
\begin{equation}
R_{\rm b}<R_{\rm lm}<R_{\rm gm}.
\end{equation}
We may interpret that the bubble radius is mostly determined by 
energy inside the bubble.

\section{Nucleation Rate and Evolution}

We begin this section by comparing the nucleation rates of the bounce 
and of gauged monopole-bubbles. In the semiclassical approximation, 
the decay rate per unit volume is estimated 
by the use of the leading exponential factor which is given by the
Euclidean action of the bubble
solution~\cite{Col}, multiplied by a prefactor determined by zero 
modes of the
fluctuation~\cite{CC}. When a continuous symmetry or a part of it is 
spontaneously broken, the number of zero modes increases and then the
tunneling rate into vacua is enhanced~\cite{Kus}. In the present 
model, the SO(3) gauge symmetry is broken
to an SO(2) symmetry and we have two bubble solutions; the relative 
decay rate
between the bounce $\phi^{a}_{\rm b}$ and the gauged monopole-bubble
$\phi^{a}_{\rm lm}$ is meaningful:
\begin{eqnarray}\label{decay2}
\frac{\Gamma_{\rm lm}}{\Gamma_{\rm b}}\sim
\left(\frac{S(\phi^{a}_{\rm lm})}{S(\phi^{a}_{\rm b})}\right)^{3/2}
\left|\frac{\det'[S''(\phi^{a}_{\rm lm})]}{\det'[S''(\phi^{a}_{\rm 
b})]}
\right|^{-1/2}\frac{\int d^3x (\phi^{a}_{\rm lm})^2}{\int d^3x
(\phi^{a}_{\rm b})^2}e^{-[S(\phi^{a}_{\rm lm})-S(\phi^{a}_{\rm b})]},
\end{eqnarray} 
where primed determinants denote infinite products over the nonzero
eigenvalues, which are assumed to be order 1 here. The value of 
the action in Eq.~(\ref{decay2}) is estimated after removing the constant
vacuum energy density at the metastable symmetric vacuum, 
that means $S(\phi=0)=0$ in Eq.~(\ref{decay2}).

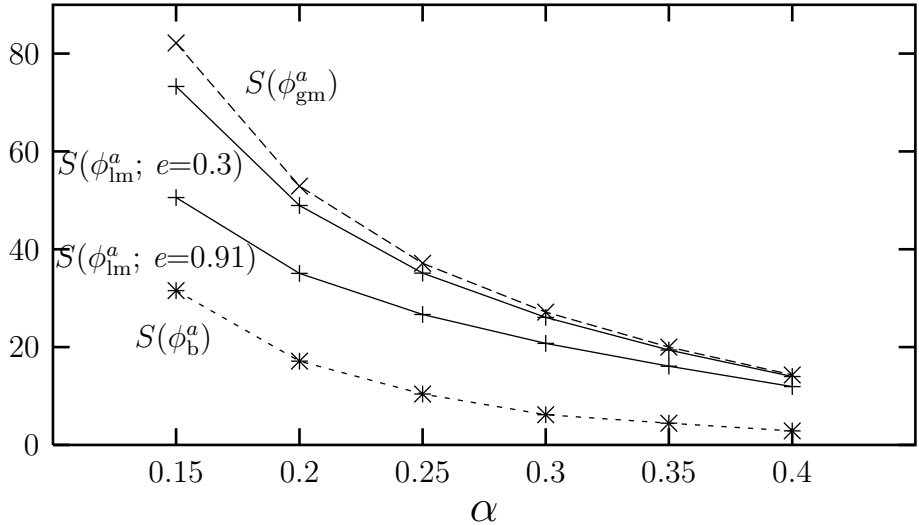
\begin{figure}
\setlength{\unitlength}{0.1bp}
\begin{picture}(3600,2160)(0,0)
\put(1925,150){\Large\makebox(0,0){$\alpha$}}
\put(3086,300){\makebox(0,0){0.4}}
\put(2621,300){\makebox(0,0){0.35}}
\put(2157,300){\makebox(0,0){0.3}}
\put(1693,300){\makebox(0,0){0.25}}
\put(1229,300){\makebox(0,0){0.2}}
\put(764,300){\makebox(0,0){0.15}}
\put(250,1876){\makebox(0,0)[r]{80}}
\put(250,1507){\makebox(0,0)[r]{60}}
\put(250,1138){\makebox(0,0)[r]{40}}
\put(250,769){\makebox(0,0)[r]{20}}
\put(250,400){\makebox(0,0)[r]{0}}
\put(1200,1750){\makebox(0,0){$S(\phi^{a}_{\rm{gm}})$}}
\put(1020,1450){\makebox(0,0)[r]{$S(\phi^{a}_{\rm{lm}}$;~{\it e}=0.3)}}
\put(1070,1100){\makebox(0,0)[r]{$S(\phi^{a}_{\rm{lm}}$;~{\it e}=0.91)}}
\put(900,800){\makebox(0,0)[r]{$S(\phi^{a}_{\rm{b}})$}}
\end{picture}
\caption{Plots of Euclidean actions $S/v\beta$ versus $\alpha$ for various 
bubble solutions. The dashed line corresponds to a global monopole-bubble 
($e=0$), two solid lines to gauged monopole-bubbles of $e=0.3$ and $0.91$, and 
the dotted line to an ordinary bounce. Here $\lambda$ is set to be 1.}
\label{fig3}
\end{figure}

In order to see the leading effect, we plot in Fig.~\ref{fig3}
the values of the Euclidean actions
for a bounce, gauged monopole-bubbles ($e=0.3$ and $0.91$), and global
monopole-bubble for $e=0$. As expected, the bounce has always
the minimum action irrespective of the thickness of the bubble wall 
(or equivalently the parameter $\alpha$) (see Fig.~\ref{fig3}) so that 
it forms a dominant decay channel (see Table~\ref{table1}). The difference 
of the Euclidean actions is almost a constant for a given gauge coupling $e$,
and it is roughly the monopole mass multiplied by inverse 
temperature. The values of the Euclidean action decrease as the gauge
coupling increases.
If we recall the mass formula $4\pi v/e$ of BPS monopole of unit 
winding~\cite{Bog}, one may easily notice that this gap decreases in 
the strong coupling limit. This behavior can easily be understood by rescaling 
the variables to the dimensionless ones: $\tilde{t}_{E}=evt_{E}$, $\rho=evr$, 
$\phi=vh$, $\tilde{V}=\tilde{\lambda}(h^{2}+\alpha)(h^{2}-1)^{2}$, and
$\tilde{\lambda}=\lambda/e^{2}$. Then the action is rewritten as
$S\sim(4\pi v\beta/e)\times ({\rm dimensionless}\;{\rm energy})$, 
which reflects the decrease of the monopole mass for strong gauge coupling.
If we take into account the enhancement due to the prefactor in Eq.~(\ref{decay2}),
we may obtain even an unbelievable relative decay rate in the high temperature
limit, e.g., $\Gamma_{\rm m}/\Gamma_{\rm b} > 1$ when $v\beta=0.1$ and $\alpha=0.4$.
However, since the transition pattern becomes weakly first-order at such high
temperature, the above result does not imply a possibility that the decay through
a monopole-bubble is dominant decay channel.
Instead, it means that in a high temperature ($v\beta\sim 1$)
the relative production rate is considerable for thick-wall bubbles
($\alpha \approx 0.4$) with strong gauge coupling ($e \sim 1$)
(see Table~\ref{table1}). For
reference we give the values of the Euclidean action for the global 
monopole-bubbles when $e=0$, which are always larger than those of the gauged 
monopole-bubbles because of their long energy tail proportional to the radius
of the global monopole-bubble.

\begin{table}
\begin{center}{
\begin{tabular}{c| c c c} \hline\hline
$\alpha$ & 0.2 & 0.3 & 0.4\\ \hline\hline
($\Gamma_{\rm gm}/\Gamma_{\rm b})|_{e=0}$ ~~& $1\times 10^{-15}$ & 
$6\times 10^{-9}$ & $1\times 10^{-4}$  \\ 
($\Gamma_{\rm lm}/\Gamma_{\rm b})|_{e=0.3}$ ~& $6\times 10^{-14}$ & 
$2\times 10^{-8}$ & $1\times 10^{-4}$ \\ 
($\Gamma_{\rm lm}/\Gamma_{\rm b})|_{e=0.91}$ & $4\times 10^{-8}$ & 
$2\times 10^{-6}$ & $8\times 10^{-4}$\\ \hline
\end{tabular}
}
\end{center}
\caption{The values of the relative decay rates,
$\Gamma_{\rm gm}/\Gamma_{\rm b}$ and $\Gamma_{\rm lm}/\Gamma_{\rm b}$,
for various $\alpha$ and $v\beta$ with $e=1$, $\lambda=1$,
and $v\beta = 1$ when we set the ratio of determinant factors to be 1.} 
\label{table1}
\end{table}

Completion of the first-order phase transition is achieved by the 
growth and percolation of nucleated bubbles. The actual process is
complicated because our O(3) bounces and gauged monopole-bubbles are 
generated at high temperature. However, one simple but probably correct way 
is to analyze the time-dependent field equations with appropriate initial 
configurations. The obtained Euclidean solution is 
time-independent both in Euclidean spacetime and in Lorentzian 
spacetime, and it therefore does not evolve in itself. Hence we give 
small radial fluctuations to the static solution as an initial 
configuration, and solve the time-dependent field equations 
numerically. If the bubble radius is smaller than the critical radius, both the 
bounce and the gauged monopole-bubble collapse; so does the magnetic monopole 
at its center. On the other hand, as Fig.~\ref{fig4} shows, if it is larger than 
the critical radius, the bubble wall starts to grow and the magnetic 
monopole remains to be 
stable with small damped oscillation. The thick wall of the 
monopole-bubble becomes a thin wall. We also find the the velocity of 
the wall approaches the light velocity. Of course, inclusion of radiation 
is needed to obtain the bubble dynamics more precisely. Here we just give
a short comment on this topic: since we have three massless gauge bosons 
outside the bubble wall and only one photon inside it, the expected pressure
difference may decrease the terminal velocity of the wall.

\begin{figure}
\setlength{\unitlength}{0.1bp}
\begin{picture}(3600,2160)(0,0)
\put(3137,1747){\makebox(0,0)[r]{$vt=4$}}
\put(3137,1847){\makebox(0,0)[r]{$vt=2$}}
\put(3137,1947){\makebox(0,0)[r]{$vt=0$}}
\put(2000,150){\makebox(0,0){\Large$vr$}}
\put(100,1230){%
\makebox(0,0)[b]{\Large\shortstack{$\phi/v$}}%
}
\put(3550,300){\makebox(0,0){10}}
\put(2930,300){\makebox(0,0){8}}
\put(2310,300){\makebox(0,0){6}}
\put(1690,300){\makebox(0,0){4}}
\put(1070,300){\makebox(0,0){2}}
\put(450,300){\makebox(0,0){0}}
\put(400,2060){\makebox(0,0)[r]{1.2}}
\put(400,1783){\makebox(0,0)[r]{1}}
\put(400,1507){\makebox(0,0)[r]{0.8}}
\put(400,1230){\makebox(0,0)[r]{0.6}}
\put(400,953){\makebox(0,0)[r]{0.4}}
\put(400,677){\makebox(0,0)[r]{0.2}}
\put(400,400){\makebox(0,0)[r]{0}}
\end{picture}

\begin{center}{(a)}
\end{center}
\setlength{\unitlength}{0.1bp}
\begin{picture}(3600,2160)(0,0)
\put(3137,1547){\makebox(0,0)[r]{$vt=4$}}
\put(3137,1647){\makebox(0,0)[r]{$vt=2$}}
\put(3137,1747){\makebox(0,0)[r]{$vt=0$}}
\put(2025,150){\makebox(0,0){\Large$vr$}}
\put(100,1230){%
\makebox(0,0)[b]{\Large\shortstack{$K$}}%
}
\put(3550,300){\makebox(0,0){10}}
\put(2940,300){\makebox(0,0){8}}
\put(2330,300){\makebox(0,0){6}}
\put(1720,300){\makebox(0,0){4}}
\put(1110,300){\makebox(0,0){2}}
\put(500,300){\makebox(0,0){0}}
\put(450,2060){\makebox(0,0)[r]{1}}
\put(450,1728){\makebox(0,0)[r]{0.96}}
\put(450,1396){\makebox(0,0)[r]{0.92}}
\put(450,1064){\makebox(0,0)[r]{0.88}}
\put(450,732){\makebox(0,0)[r]{0.84}}
\put(450,400){\makebox(0,0)[r]{0.8}}
\end{picture}

\begin{center}{(b)}
\end{center}
\caption{Evolution of a gauged monopole-bubble with a thick-wall: (a) scalar
amplitude $\phi/v$ and (b) gauge field $K$, where $\lambda=1$, $e=0.3$, 
and $\alpha=0.4$.}
\label{fig4}
\end{figure}

\section{Concluding Remarks}

We have seen that when a first-order phase transition in the gauge theory in which
the symmetry breaking pattern is SO(3)$\rightarrow$SO(2) is considered in the high
temperature limit; a new gauged monopole-bubble solution was found.
It is distinguished from the known Euclidean
bounce by the production of a 't Hooft-Polyakov monopole at the center of
the bubble the moment it is nucleated. The production rate of
monopole-bubbles is smaller than that of the bounce, but it is considerable
for thick-wall cases with strong gauge coupling in the limit of high 
temperature.
When the size of a nucleated bubble is larger than the critical size, the
bubble wall starts to move outward so that the magnetic monopole remains
stable in the bubble at least before bubble collision.

In a theoretical sense, the existence of such a gauged monopole-bubble solution
demonstrates clearly a possible drastic effect of the gauge field on 
bubble nucleation.
Although it is important to understand the evolution of the gauged 
monopole-bubble
through zero-modes, the stability including angular fluctuations
is left for future work. The realization of these 
monopole-bubbles in a real material is not known yet.

\section*{Acknowledgments}

{The authors would like to thank Kyoungtae Kimm for discussions.
Numerical Computation of this work was carried out at the Yukawa Institute Computer Facility. 
Y.K. wishes to acknowledge the financial support of the Korea Research
Foundation made in the program year of 1997. 
N. S. was supported by JSPS Research Fellowships for Young Scientist. This work was supported partially by the
Grant-in-Aid for Scientific Research Fund of the Ministry of Education, 
Science, Sports and Culture (No.\ 9702603 and No.\ 09740334) and by the 
Waseda University Grant for Special Research Projects.}


\end{document}